\documentclass[10pt,conference]{IEEEtran}

\IEEEoverridecommandlockouts
\usepackage{cite}
\usepackage{enumitem,amsmath,amssymb,amsfonts}
\usepackage{algorithmic}
\usepackage{graphicx}
\usepackage{textcomp}
\usepackage{xcolor}
\usepackage[hyphens]{url}
\usepackage{hyperref}
\usepackage{booktabs}
\usepackage{multirow}
\usepackage{multicol}
\usepackage{fancyhdr,lipsum}
\usepackage[T1]{fontenc}
\usepackage{tabularx}
\usepackage{graphicx}
\usepackage{tikz}
\usepackage{subcaption}
\usepackage{makecell}
\usepackage{upquote}
\usepackage{balance}
\usepackage[most]{tcolorbox}

\newlist{todolist}{itemize}{2}
\setlist[todolist]{label=$\square$}

\begin{document}

\title{Trustworthy and Synergistic Artificial Intelligence for Software Engineering: Vision and Roadmaps}

\author{\IEEEauthorblockN{David Lo}
\IEEEauthorblockA{School of Computing and Information Systems,\\ Singapore Management University,\\Singapore\\
Email:\:davidlo@smu.edu.sg
}}

\maketitle

\begin{abstract}

For decades, much software engineering research has been dedicated to devising automated solutions aimed at enhancing developer productivity and elevating software quality. The past two decades have witnessed an unparalleled surge in the development of intelligent solutions tailored for software engineering tasks. This momentum established the Artificial Intelligence for Software Engineering (AI4SE) area, which has swiftly become one of the most active and popular areas within the software engineering field.

This Future of Software Engineering (FoSE) paper navigates through several focal points. It commences with a succinct introduction and history of AI4SE. Thereafter, it underscores the core challenges inherent to AI4SE, particularly highlighting the need to realize trustworthy and synergistic AI4SE. Progressing, the paper paints a vision for the potential leaps achievable if AI4SE's key challenges are surmounted, suggesting a transition toward Software Engineering 2.0. Two strategic roadmaps are then laid out: one centered on realizing trustworthy AI4SE, and the other on fostering synergistic AI4SE. While this paper may not serve as a conclusive guide, its intent is to catalyze further progress. The ultimate aspiration is to position AI4SE as a linchpin in redefining the horizons of software engineering, propelling us toward Software Engineering 2.0.\vspace{0.2cm}

\end{abstract}


\thispagestyle{plain}
\pagestyle{plain}

\section{Introduction and Brief History of AI4SE}\label{sec:introduction}
%``The road to the future runs through the past'' -- Robert Webber

\vspace{0.1cm}``{\em Study the past if you would define the future.}'' -- Confucius\vspace{0.2cm}

%``{\em Without the past, there is no future}'' -- Anonymous\vspace{0.2cm}

%-- ``{\em There is no future without a past, because what is to be cannot be imagined except as a form of repetition}''

Software engineering encompasses many tasks spanning the various phases of software development, from requirement gathering and design to coding, testing, and deployment. To boost developer productivity and ensure high-quality software, extensive research in software engineering has aimed to automate some of these manual tasks. While initial automation efforts centered around the development of program analysis methods, e.g., linters~\cite{Johnson77}, model checkers~\cite{ClarkeE81,QueilleS82},  fuzzers~\cite{MillerFS90}, etc., the past two decades have witnessed a rapid rise in the design and deployment of AI-powered solutions to assist software practitioners in their tasks.

\begin{figure}[t]
    \centering
    \includegraphics[width=3.4in]{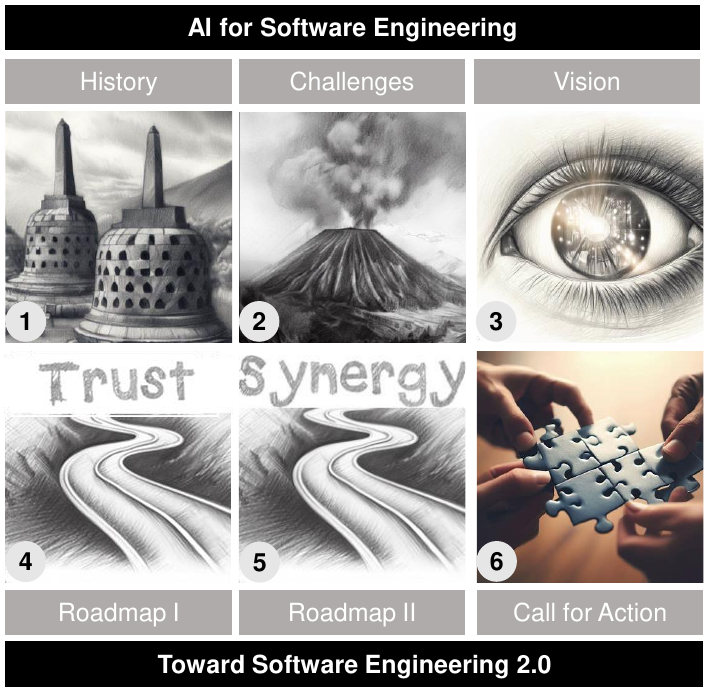}
    \caption{Overview of the Paper with Its Six Sections}
    \label{fig:overview}
\end{figure}

AI-powered solutions have been employed to analyze a myriad of software artifacts, both products and by-products of software engineering activities. These artifacts encompass source code, execution traces, bug reports, and posts on question-and-answer sites, among others. A variety of AI methods underpin the development of these assistive solutions. In this paper, the term ``AI'' is employed in a broad sense, encompassing a range of techniques from data mining and information retrieval to meta-heuristics, natural language processing, and machine learning. Today, the boundaries between these fields are becoming increasingly blurred, as many emerging techniques span across them.

As illustrated in Fig.~\ref{fig:overview}, this paper first describes the history and key challenges of AI for Software Engineering (AI4SE) in Sections~\ref{sec:introduction} and~\ref{sec:challenges}, respectively. In then describes a vision for AI4SE in Section~\ref{sec:vision}. Next, it highlights two roadmaps toward trustworthy and synergistic AI4SE in Sections~\ref{sec:roadmap_trust} and~\ref{sec:roadmap_synergy}, respectively. The paper concludes by providing a summary and a call for action in Section~\ref{sec:conclusion}.

The remainder of this section offers a concise overview of AI4SE's history after  ``AI winters'' ended in the mid-2000s.\footnote{``Scientists and writers describe the dashed hopes of the period from the 1970s until the mid-2000s as a series of `AI winters.'''~\cite{Bengio16}} Even focusing from the mid-2000s to today, there are too many studies to cover. Thus, this section does not attempt to provide comprehensive coverage. For more coverage of the history of AI4SE, please refer to other reviews and surveys. An example review of the emergence of AI4SE after the ``AI winters'' is by Xie et al.~\cite{XieTLL09} on data mining for software engineering. The article described the applications of four AI techniques -- frequent pattern mining, pattern matching, clustering, and classification -- to analyze software engineering artifacts such as execution traces, source code, and bug reports, which can be represented in forms like sequences, graphs, and text. Notably, it described specialized AI techniques adept at inferring software's formal specifications (e.g.,~\cite{LoK06}), synthesizing bug signatures (e.g.,~\cite{ChengLZWY09}), and identifying duplicate bug reports (e.g.,~\cite{WangZXAS08}). Beyond this review, there are others covering various AI4SE topics, e.g.,~\cite{BinkleyL10,RobillardBKMR13}.\footnote{For a review on studies in the intersection of AI and software engineering research before the end of ``AI winters'', see, for instance,~\cite{PARTRIDGE1988146}.}

%One of the early reviews on the topic of AI for Software Engineering (AI4SE) is a 2009 IEEE Software cover article ``Data Mining for Software Engineering''~\cite{XieTLL09}. The article highlights four AI methods, i.e. frequent pattern mining, pattern matching, clustering, and classification, which can be applied to analyze sequences, graphs, and text that correspond to software engineering artifacts (execution traces, source code, and bug reports). The review article highlighted a number of early works that use data mining methods to help in requirement engineering, program comprehension, verification, debugging, and software maintenance and evolution. Specifically, it highlighted specialized AI methods that can infer formal specifications of software~\cite{LoK06}, infer bug signatures~\cite{ChengLZWY09}, and detect duplicate bug reports~\cite{WangZXAS08}. 

Over the past two decades, significant efforts have been made to solidify AI4SE as a recognized research area within the software engineering field. A prominent one is the Mining Software Repositories (MSR) conference series, inaugurated as a workshop at the 26th ACM/IEEE International Conference on Software Engineering (ICSE 2004). This workshop ``brought together researchers and practitioners in
order to consider methods that use data stored in software repositories (such as source control systems, defect tracking systems, and archived project communications) to further understanding of software development practices.''~\cite{HassanHM05} Another prominent conference series is the Symposium on Search Based Software Engineering (SSBSE), which started in 2009~\cite{ssbse2009}, and serves as a forum that focuses on solving software engineering tasks by formulating them as optimization or search problems. 

A number of AI4SE workshops have also made significant contributions. One that ran for many years (2012 to 2018) was the International Workshop on Software Mining (SoftMine). SoftMine ``facilitated researchers who are interested in mining various types of software-related data and in applying data mining techniques to support software engineering tasks.''~\cite{LiZLL15} By being hosted at both SE and AI conferences -- including the IEEE/ACM International Conference on Automated Software Engineering (ASE) and the ACM SIGKDD Conference on Knowledge Discovery and Data Mining (KDD) -- SoftMine bridged the gap between the SE and AI communities. Additionally, numerous AI4SE tutorials and summer/winter schools have been organized over the years.

Today, these community efforts have borne fruit, evidenced by the prominence of AI4SE at leading software engineering conferences. It is noteworthy that at the 45th ACM/IEEE International Conference on Software Engineering (ICSE 2023), ``Artificial Intelligence and Software Engineering'' and ``Software Analytics'' were not only among the conference's primary seven areas but also among its most popular.

There have been multiple ``waves'' that positively influence the area of AI4SE:

\vspace{0.2cm}
\noindent{\em Wave 1 (Big Software Engineering Data)} 
\vspace{0.2cm}

By the late 2000s, an increasing volume of open-source data became accessible to researchers, marking a significant shift from the time when most software artifacts remained confined within corporate boundaries. Notably, 2008 witnessed the launch of platforms like GitHub and Stack Overflow. The repositories on GitHub have expanded consistently over the years, and Stack Overflow's post count has shown a similar trajectory. These platforms supply vast amounts of data for AI4SE. As early as 2013, SE researchers had analyzed tens of thousands of version control and issue tracking systems from GitHub repositories~\cite{ThungBLJ13,BissyandeLJRKT13,BissyandeTLJR13}. Additionally, these platforms have presented unique challenges that demand AI4SE solutions. For example, the high number of posts in Stack Overflow requires the design of solutions that can empower the community to better browse~\cite{XiaLWZ13,beyer2015synonym}, locate~\cite{CalefatoLN19,gao2020technical}, maintain~\cite{zhang2019empirical,zhang2021study}, and comprehend~\cite{XuXXL17,YangXTSZ0ZSHH022} answers to software engineering questions.

\vspace{0.2cm}
\noindent{\em Wave 2 (Deep Learning for Software Engineering)}
\vspace{0.2cm}

By the 2010s, deep learning has gained much traction across various domains, beyond computer vision. This evolution significantly influenced the AI4SE landscape. Initial forays into integrating deep learning with software engineering include constructing deep learning solutions for defect prediction~\cite{yang2015deep} and code suggestion~\cite{WhiteVVP15}. Subsequent research expanded upon various deep learning architectures, such as the Recurrent Neural Network and Transformer, aiming to automate a wider range of software engineering tasks. Moreover, considerable effort went into devising methods to learn effective distributed representations of diverse software artifacts, exemplified by works like~\cite{AlonZLY19,HoangK0L20}. Comprehensive surveys on this topic were written by Yang et al.\cite{YangXLG22} and Watson et al.\cite{WatsonCNMP22}.

While there have been notable successes in marrying deep learning and software engineering, there are documented instances of limitations. Some studies, such as~\cite{MajumderBBFM18,LiuXHLXW18,ZengZZZ21}, demonstrate that, in certain situations and tasks, the efficacy of simpler techniques may be comparable to, or even surpass, deep learning models. Moreover, concerns about the generalizability of representations derived via deep learning are also highlighted by some studies such as~\cite{KangB019}.

%In the 2010s, deep learning has became very popular in many areas beyond computer vision. This impacted the area of AI4SE. Early work on deep learning for software engineering  include its application for defect prediction~\cite{yang2015deep} and code suggestion~\cite{WhiteVVP15}. Since then, there have been a wave of research that apply and build upon different deep learning architectures, e.g., Recurrent Neural Network, Transformer, etc. to automate many software engineering tasks. There has also been much research on the design of methods that can learn good distributed representations of different software artifacts, e.g.,~\cite{AlonZLY19,HoangK0L20}. Two comprehensive surveys about this has been written by Yang et al.~\cite{YangXLG22} and Watson et al.~\cite{WatsonCNMP22}. There have been studies that highlight both success and failure cases of deep learning for software engineering. The latter types of studies highlight that at times simpler approach can perform the same or better than deep learning, e.g.,~\cite{MajumderBBFM08}. It also highlights the generalizability problem of representations learned by deep learning, e.g.,~\cite{KangB019}.

\vspace{0.2cm}
\noindent{\em Wave 3 (Large Language Model or Foundation Model for Software Engineering)}
\vspace{0.2cm}

This wave continues from Wave 2. In 2018, Google introduced the Bidirectional Encoder Representations from Transformers (BERT), which underwent pre-training on the Toronto Book Corpus and English Wikipedia~\cite{DevlinCLT19}. Often considered the first Large Language Model (LLM) or Foundation Model (FM)\footnote{The terms ``Large Language Model'' and ``Foundation Model'' are currently frequently used synonymously. However, a Foundation Model can encompass more than just a Large Language Model. For instance, it can support multiple modalities, including images. Since most software engineering studies today utilize Large Language Models as their Foundational Models, this paper predominantly uses the term ``Large Language Model.''}, BERT can significantly reduce the necessity for large amounts of high-quality labeled data for downstream tasks. By 2020, researchers had applied BERT to automate software engineering tasks~\cite{ZhangXTH0J20,BiswasKPV20}. That same year, Microsoft unveiled CodeBERT, which is a variant of BERT pre-trained on a corpus containing a mixture of source code and natural language text~\cite{FengGTDFGS0LJZ20}. After BERT, ever larger LLMs have been developed, with models like GPT-4~\cite{abs-2303-08774} currently leading the way. These LLMs have been leveraged to automate many software engineering tasks~\cite{hou2023large}. Interestingly, many years before the development of LLMs, Hindle et al.~\cite{HindleBSGD12} have underscored the inherent naturalness of software and the prospective utility of language models in automating software engineering tasks.

The most recent wave has significantly advanced the adoption of AI4SE solutions among practitioners. Today, solutions like GitHub Copilot, Amazon CodeWhisperer, and OpenAI ChatGPT, built upon LLMs, are used by many professional and aspiring software practitioners. While, as of this paper's writing, their primary applications have been in coding, it is easy to envision their expansion into a broader spectrum of software engineering tasks in the near future, such as design, requirement elicitation, verification, and bug report management, among others.

\section{Key Challenges of AI4SE: Trust and Synergy}\label{sec:challenges}
\vspace{0.1cm}``{\em Victory comes from finding opportunities in problems}''

-- Sun Tzu \vspace{0.2cm}

While there has been a surge in AI4SE research and its widespread adoption, numerous challenges remain, offering ample opportunities for future research. Many of these challenges can be put into two broad categories: ensuring {\bf trustworthy AI4SE} and promoting {\bf synergistic AI4SE}. If AI4SE solutions are not trusted by practitioners, they will not be adopted. Trust is dynamic; AI4SE solutions need to maintain practitioners' confidence in them over time. Moreover, an effective AI4SE solution should not only be trustworthy but also synergize seamlessly with practitioners. If not, such AI4SE solutions risk becoming obstacles rather than facilitators. While trust and synergy are interconnected concepts, each has its unique characteristics.

\subsection{Need for Trust}

In 2015, a study involving 512 Microsoft practitioners was carried out to assess their perceptions on the relevance of software engineering research~\cite{LoNZ15}. The goal was to identify potential gaps between academic research and its practical application. The results underscored concerns from practitioners, including those revolving on trust. For instance, one respondent stated ``It seems that there could be potentially disastrous results if the
automation does not [do things] correctly.'' A follow-up study in 2016, involving 386 practitioners from more than 30 countries across 5 continents, highlighted similar findings~\cite{KochharXLL16}. For example, one respondent stated, ``I doubt any automated software can explain the reason for things $\ldots$'', highlighting a reason behind the lack of trust.

Fast forward to 2023, and although AI4SE research has undoubtedly advanced since 2015, challenges persist. A 2022 study revealed that many code snippets produced by GitHub Copilot contain security vulnerabilities~\cite{PearceA0DK22}. Similarly, a 2023 article pointed out that code generated by ChatGPT often had compilation and runtime errors, especially when applied to newer programming tasks that might not have been present in its training data~\cite{Liu23}.  Recent news articles further emphasize these concerns, such as ``Friend or foe: Can computer coders trust ChatGPT?''~\cite{McManus23} and ``ChatGPT creates mostly insecure code, but won't tell you unless you ask''~\cite{Claburn23}. There are also concerns about significant variations in ChatGPT's efficacy over time~\cite{abs-2307-09009}, including in the code generation task. Such inconsistencies can undermine trust as practitioners seek stability in the efficacy of AI4SE solutions~\cite{ChenSASS020}.

If these trust issues are not adequately addressed, the current enthusiasm surrounding AI4SE can possibly diminish, reminiscent of the declining interest in AI experienced during the ``AI winters.''~\cite{mccorduck2004}. Another side of the trust spectrum is over-reliance. Novices may mistakenly place excessive trust in these AI4SE solutions, expecting flawless results, while remaining oblivious to their inherent limitations, which can prove detrimental.

\subsection{Need for Synergy}

Synergy is typically defined as the collaboration of two or more entities to create an outcome that exceeds the sum of their individual contributions. In the context of AI4SE solutions, there are two primary usage scenarios: {\tt 1:1}, where a single software practitioner interacts with an AI4SE solution, and {\tt N:M}, involving multiple software practitioners collaborating with multiple AI4SE solutions. Presently, the majority of research and existing AI4SE solutions focus on the {\tt 1:1} scenario, which is intuitively more straightforward than the {\tt N:M} scenario. However, even within this simpler setting, there is no guarantee that software practitioners and AI4SE solutions will achieve seamless synergy. 

\vspace{0.3cm}
\noindent{\em Interlink between synergy and trust}
\vspace{0.3cm}

Synergy and trust are intrinsically linked; it is challenging for two entities to work together seamlessly without trust. As an example, Parnin and Orso conducted a controlled experiment showing that while fault localization solutions\footnote{A fault localization solution produces a ranked list of program locations that are likely to be faulty~\cite{JonesH05,ZhangLMYXL23}. They typically take as input a collection of program spectra describing program locations that are executed by failing and successful test cases.} can achieve favorable results by certain metrics, they do not necessarily expedite human debugging processes~\cite{ParninO11}. Another research, although not centered on AI4SE, underscores that many professionals avoid static analysis solutions due to their frequent false positives, among other reasons~\cite{JohnsonSMB13,ChristakisB16}. These studies underscore the ``boy who cried wolf'' phenomenon: when AI4SE solutions (or any automated solution for that matter) repeatedly produce unreliable results, software practitioners become skeptical and may disengage, preventing any synergy.

Once trust is firmly established, AI4SE solutions and practitioners can potentially harmonize in ways that yield tangible benefits. For instance, for fault localization, if an AI4SE solution consistently presents accurate outcomes within the top-5 or top-10 results, Xia et al. found that practitioners can gain a significant performance boost~\cite{XiaBLL16}. But the bar for adoption can be steep: Kochhar et al. noted that for 90\% of practitioners to adopt fault localization solutions, these solutions must deliver accurate results within the top-5 positions at least 90\% of the time~\cite{KochharXLL16}. Achieving such a goal is undeniably challenging.

This narrative emphasizes that synergy and trust are closely intertwined, with trust being influenced by the efficacy of AI4SE solutions, specifically their capability and likelihood to yield accurate results. Although slight enhancements in efficacy may not immediately foster trust, there exists a critical threshold that, once exceeded, can serve as a tipping point for both trust and synergy. Therefore, as a community, it is essential to continually push for greater efficacy, even if immediate gains in trust and synergy are not readily visible.

\vspace{0.3cm}\noindent{\em Synergy beyond trust} \vspace{0.3cm}

While synergy undoubtedly involves trust, it extends beyond that singular concept. Mere trust does not ensure that the collaboration between two entities will yield outcomes surpassing their separate contributions. Some barriers stand in the way of achieving synergy between software practitioners and AI4SE solutions, including the following:

\vspace{0.2cm}\noindent{\em  Piscem natare doces}: In the survey of Microsoft practitioners mentioned earlier~\cite{LoNZ15}, several participants pointed out that they deemed certain research unnecessary, as the resultant solutions were not seen as essential. This sentiment was particularly more pronounced among experts; the study found that as experience increased, participants were more critical and considered more studies as unimportant as well as fewer studies as essential. This observation was statistically significant with a p-value of 0.01.

\vspace{0.2cm}\noindent Disrupting the ``flow'': Software practitioners are most effective when they are in a state of ``flow'', a concept described as ``a state in which people are so involved in an activity that nothing else seems to matter''~\cite{csikszentmihalyi1990flow}. This state has been proven crucial for the productivity of software practitioners~\cite{MeyerFMZ14,ForsgrenSMZHB21}. If AI4SE solutions are introduced inappropriately or at inopportune moments, software practitioners may feel disrupted, much like the annoyance users felt with Microsoft's Clippy, which was deemed ``annoying, impolite, and disruptive of a user's workflow''~\cite{veletsianos2007cognitive}. 

\vspace{0.2cm}\noindent{Resistance to change}: People often display an aversion to modifying their established routines. Previous research has highlighted software practitioners' hesitancy to embrace new processes~\cite{AnastassiuS20} or technologies~\cite{AmorimVPGP19}. Merely introducing an AI4SE solution does not ensure immediate adoption and endorsement, especially if it disrupts familiar practices. Such resistance can arise if AI4SE solutions are not smoothly integrated into the technological environments software practitioners are accustomed to. For instance, many AI4SE solutions have not been integrated into popular IDEs or issue-tracking systems, hampering their adoption. Another possible situation where resistance may happen is when employing an AI4SE solution may necessitate practitioners to adopt new procedures. For example, to leverage a fault localization solution, practitioners are prompted to consult a ranked list of potentially buggy program locations -- a step they may not practice before the introduction of such a solution. The need for a change in {\em modus operandi} may pose a certain resistance that needs to be effectively managed (c.f.,~\cite{lawrence1969deal}).

\vspace{0.2cm}\noindent{Differences in abstraction levels}: Practitioners often consider overarching goals, which may involve a workflow of many tasks, each further broken down into micro-tasks. For example, the DevOps workflow consists of multiple phases, and each phase, e.g., coding, includes multiple activities, e.g., navigation, editing, comprehension, etc.~\cite{MinelliML15,XiaBLXHL18}.  In contrast, current AI4SE solutions usually target specific, narrower micro-tasks, such as fault localization, clone detection, API recommendation, code summarization, duplicate bug report detection, etc. While each of these micro-tasks is important, the lack of understanding of the overarching workflow may be a barrier to effective synergy.

\vspace{0.1cm}\noindent{Communication barriers}: The way humans communicate with each other when collaborating on tasks differs from human-AI4SE solution interactions. While humans have a wide range of communication means to collaborate with each other -- text, code, sketches~\cite{mangano2014software}, and so forth -- their communication means with AI4SE solutions are much more limited. Moreover, humans engage in multi-round exchanges~\cite{clark1991grounding}, drawing from both short~\cite{baddeley2003working} and long-term~\cite{Tulving1973EncodingSA} memories of past interactions. Many AI4SE solutions, on the other hand, operate in a single interaction mode. For instance, in fault localization~\cite{JonesH05,ZhangLMYXL23}, most solutions simply allow practitioners to provide a set of program spectra (corresponding to failing and successful test cases). The AI4SE solution then returns a list of potential faulty program locations. A similar observation can be made for many other AI4SE solutions, e.g., code search~\cite{ShiYW0Z0ZX22}, code summarization~\cite{HuLXLJ18}, etc.

%\end{itemize}

\vspace{0.2cm}These challenges can hinder effective synergy, making it difficult for practitioners to fully benefit from AI4SE solutions.

\section{Vision: Software Engineering 2.0}\label{sec:vision}
\vspace{0.1cm}``{\em Anything one can imagine, others can make real}''

-- Jules Verne \vspace{0.2cm}

What possibilities can a trustworthy and synergistic AI4SE unveil? This section paints a future shaped by trustworthy and synergistic AI4SE. Different from conventional papers that chronicle past achievements, this section chooses instead to spotlight the potential of what can be, given the necessary leaps in innovation. With trustworthy and synergistic AI4SE maturing, we are steadily advancing toward establishing a {\em symbiotic partnership between software practitioners and autonomous,  responsible, and intelligent AI4SE agents, creating a human-AI hybrid workforce}. The realization of this human-AI hybrid workforce heralds a new era of {\bf Software Engineering 2.0 (SE 2.0)}.\footnote{Software Engineering 2.0 is distinct from the concept of Software 2.0. Software 2.0 is defined as software that ``is written in much more abstract, human unfriendly language, such as the weights of a neural network. No human is involved in writing this code.''~\cite{karpathy2017software} Software Engineering 2.0 focuses on constructing autonomous, responsible, intelligent AI4SE agents that can symbiotically work with software practitioners to collaboratively build software, whether they are Software 1.0, Software 2.0, or possibly future Software X.0. Given the unique advantages and limitations of both Software 1.0 and Software 2.0, it is anticipated that future software systems will integrate both, thereby creating complex composite software systems where the need for Software Engineering 2.0 will be even more significant.}

%As Software 2.0 does not unequivocally supersede Software 1.0, in the future, we are likely to see a mixture of Softwa

%Software Engineering 2.0 applies to the engineering of any software systems, be it Software 1.0, Software 2.0, and possibly future Software X.0. It is likely that we will see a mixture of software in future systems as Software 1.0 and Software 2.0 have their own strengths and weaknesses -- Software 2.0 is not an absolute winner to Software 1.0. 

\begin{figure*}[ht]
    \centering
    \begin{tabular}{cc}
        \begin{subfigure}{0.45\linewidth}
            \includegraphics[width=\linewidth]{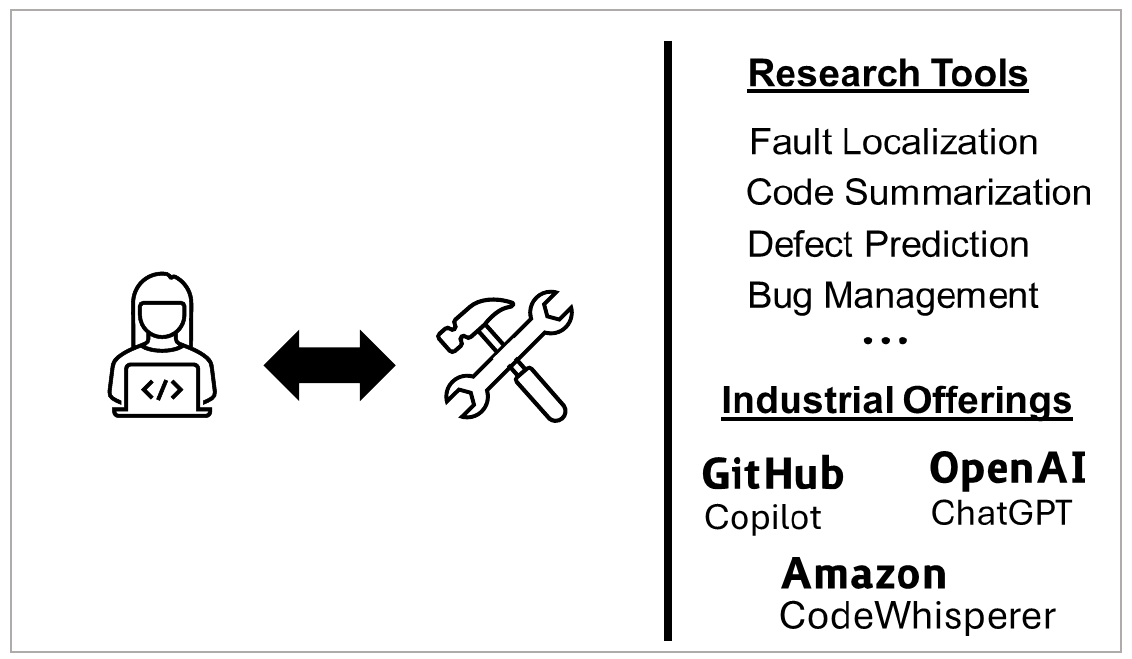}
            \caption{Current state}
        \end{subfigure} &
        \begin{subfigure}{0.45\linewidth}
            \includegraphics[width=\linewidth]{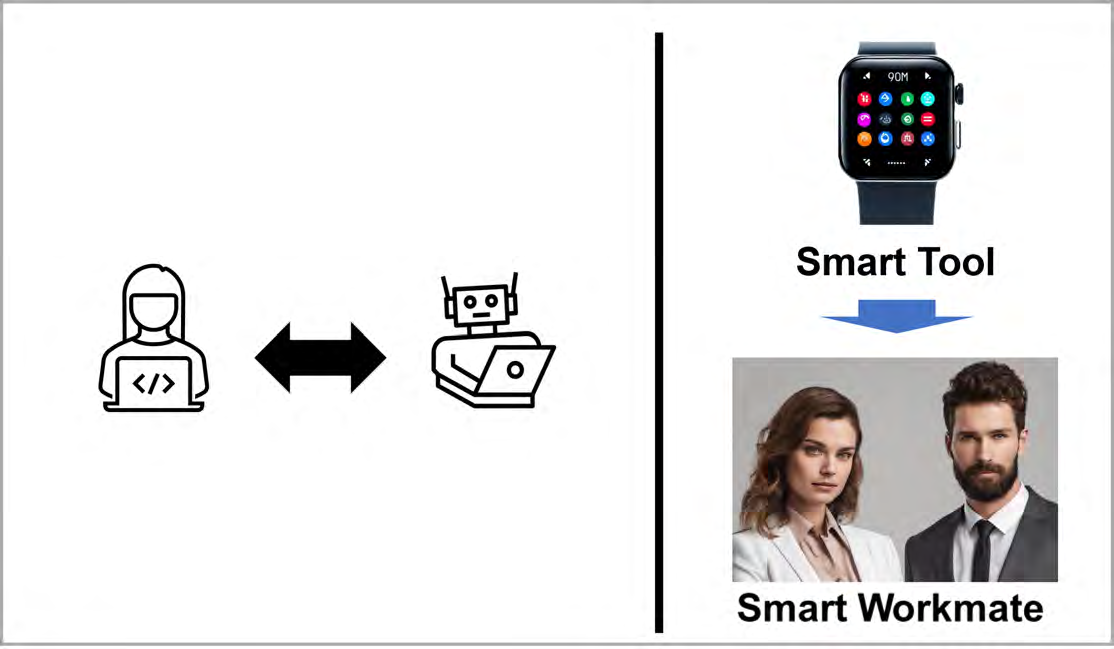}
            \caption{Smart tool to  workmate, exercising {\em responsible autonomy}}
        \end{subfigure} \\
        \begin{subfigure}{0.45\linewidth}
            \includegraphics[width=\linewidth]{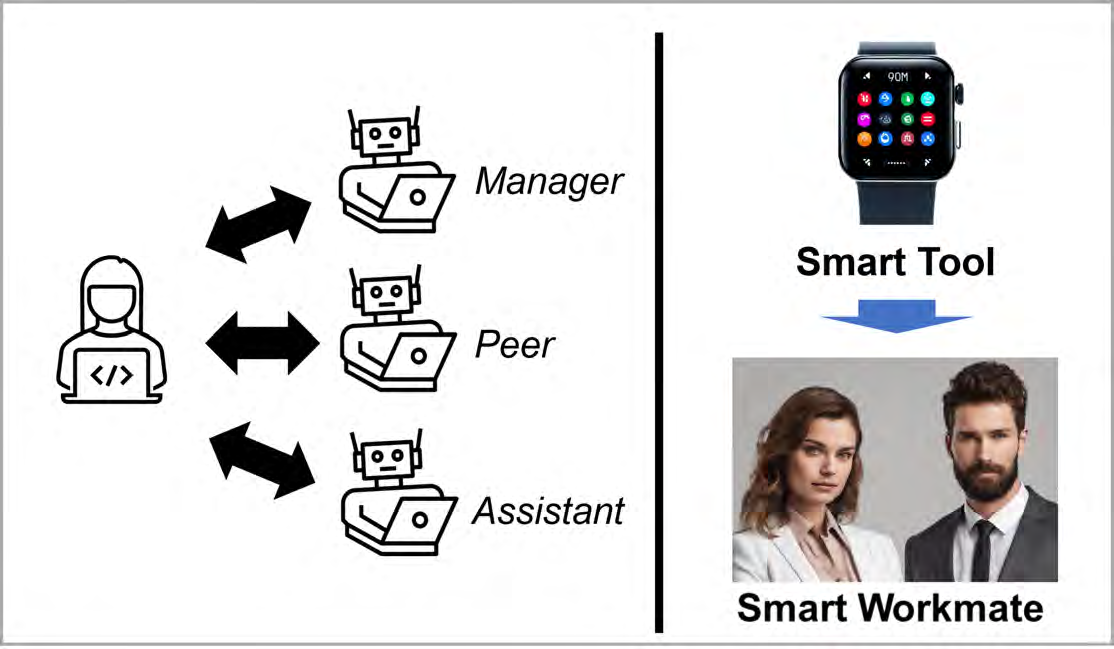}
            \caption{First-class citizens, taking different roles}
        \end{subfigure} &
        \begin{subfigure}{0.45\linewidth}
            \includegraphics[width=\linewidth]{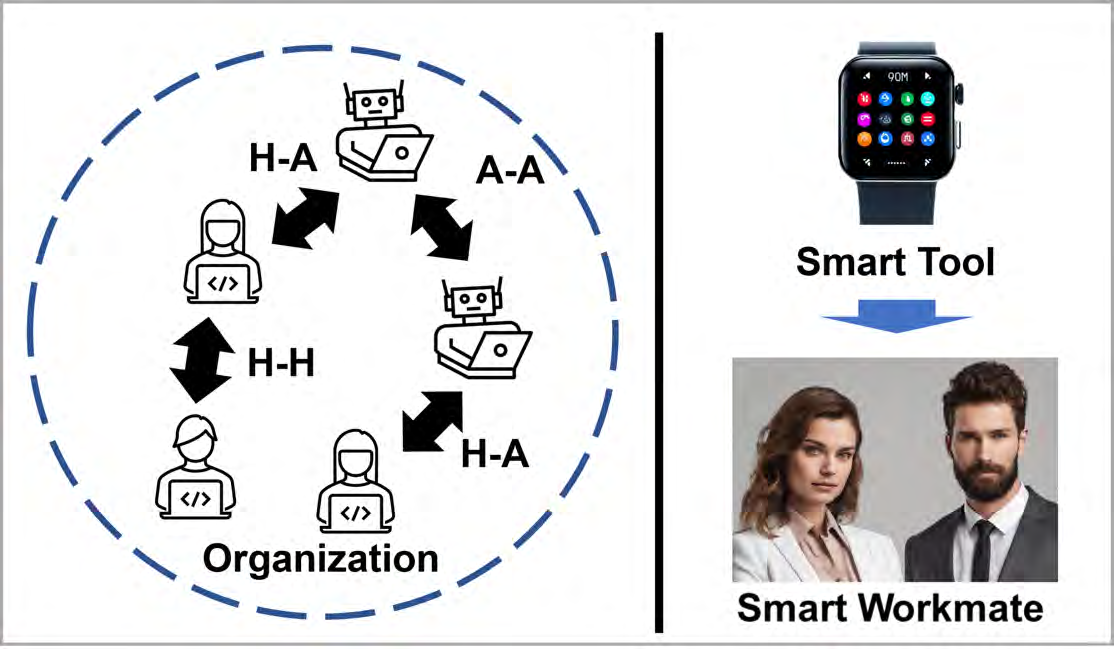}
            \caption{Able to work in a team with diverse interactions}
        \end{subfigure} \\
        \begin{subfigure}{0.45\linewidth}
            \includegraphics[width=\linewidth]{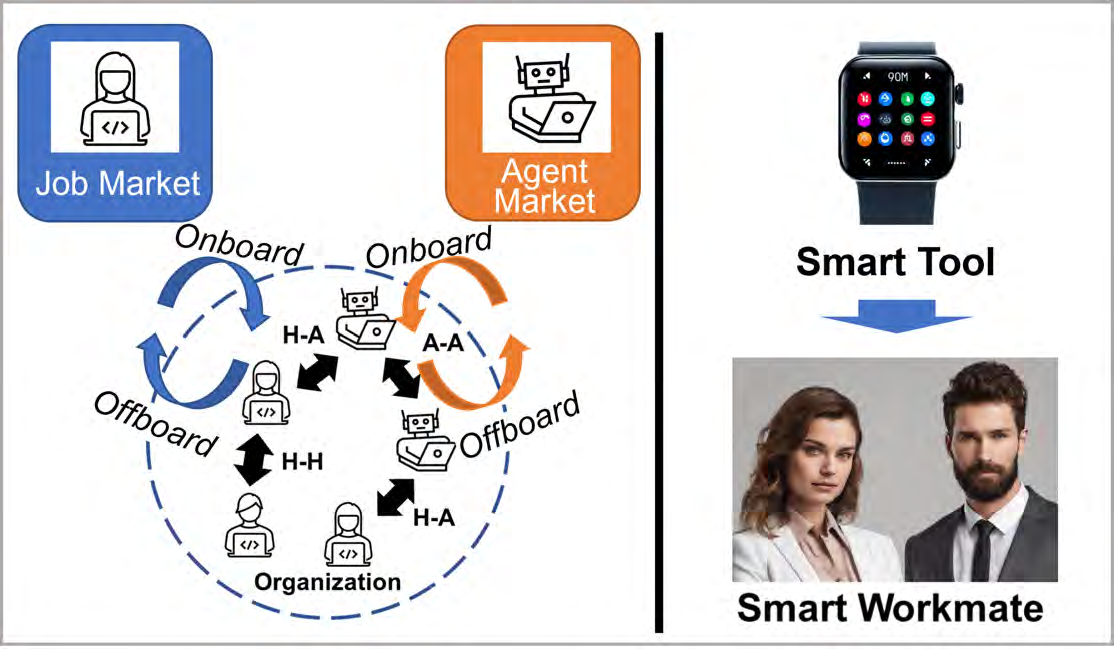}
            \caption{Able to deal with dynamic environment}
        \end{subfigure} &
        \begin{subfigure}{0.45\linewidth}
            \includegraphics[width=\linewidth]{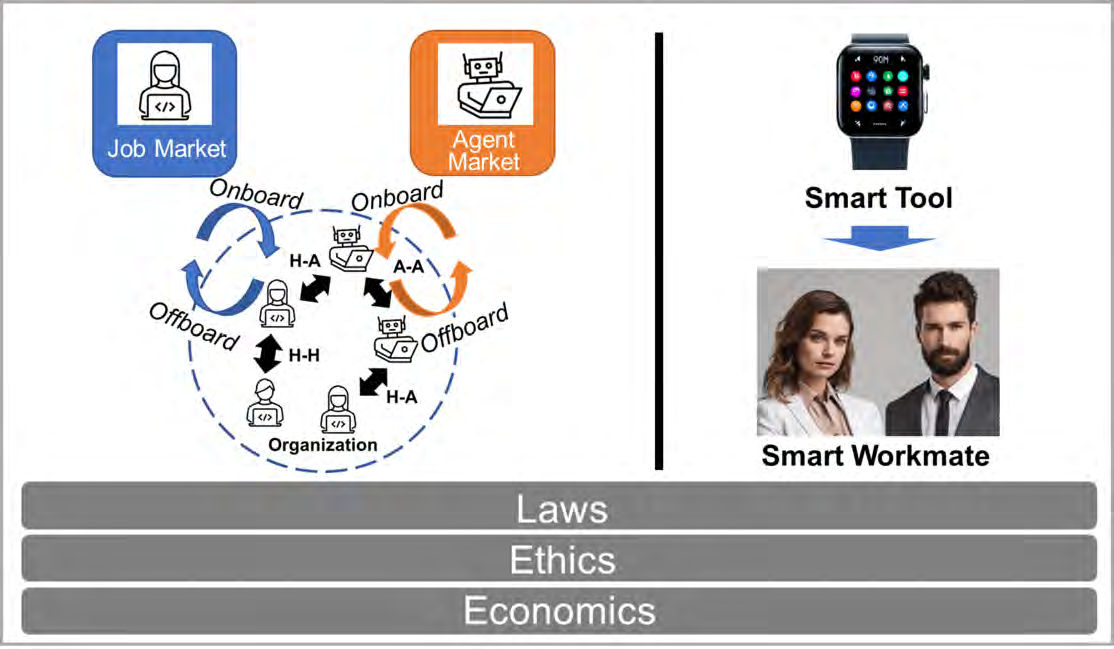}
            \caption{Require solid foundations in multiple disciplines}
        \end{subfigure} \\
    \end{tabular}
    \caption{Software Engineering 2.0 (SE 2.0) will see AI4SE solutions transitioning from smart tools to smart workmates. These intelligent agents will exercise responsible autonomy. They will be first-class citizens, taking different roles as assistants, peers, and even managers in software engineering projects. They will be well integrated in dynamic software engineering teams. There will be human-human (H-H), human-agent (H-A), and agent-agent (A-A) interactions. There will also be a new economy surrounding the AI4SE agent market as well as solid foundations in law and ethics supporting SE 2.0.}
    \label{fig:vision}
\end{figure*}

\vspace{0.2cm}
\noindent{\em Current state} 
\vspace{0.2cm}

Over the past two decades of research in AI4SE, we have witnessed the rise of AI4SE tools. As shown in Fig.~\ref{fig:vision}(a), platforms like GitHub Copilot, Amazon CodeWhisperer, and OpenAI ChatGPT, are now embraced by many professional and aspiring software practitioners. This represents a big shift from AI4SE's nascent stages when AI capabilities were limited, drawing insights only from much smaller datasets for binary predictions (such as defect prediction~\cite{MenziesGF07} and failure prediction~\cite{LoCHKS09}) or generating contents confined to a strict form or grammar (such as specification mining~\cite{LoK06}).

However, to actualize the vision of Software Engineering 2.0, there remain significant challenges to address. SE 2.0 envisions a reality that transcends merely equipping developers with AI tools for rudimentary tasks—like generating standard boilerplate code or commonplace algorithms; tasks that today's tools have already mastered. Rather, it gestures toward a broader, more transformative horizon.

\vspace{0.2cm}
\noindent{\em From smart tools to smart workmates} 
\vspace{0.2cm}

As research in trustworthy and synergistic AI4SE advances, AI4SE tools will evolve from mere smart tools to smart workmates — see Fig.~\ref{fig:vision}(b). The defining trait that distinguishes a smart workmate from a smart tool is {\em responsible autonomy} -- something we expect from a human colleague in software development. When a tool is trustworthy and can synergize well with practitioners, it can operate with increased {\em autonomy}, with practitioners confident in its ability to {\em responsibly} execute tasks. This is a big transition analogous to a transformation from a smartwatch to a colleague, reminiscent of the android Data from the Star Trek series. 

Moreover, currently, AI4SE tools predominantly function as assistants. As these tools mature further, they stand to be recognized as {\em first-class citizens} within software development processes. As first-class citizens, these tools will evolve into smart workmates that can assume a broader array of roles. They can act as peers, aiding in the development of software modules with limited supervision, or even adopt managerial capacities, performing work planning and coordination tasks -- see Fig.~\ref{fig:vision}(c). 

Furthermore, AI4SE workmates, acting as intelligent agents, will not be limited to collaborating with just one individual. Instead, they will become integral members of teams. Picture a blended team of software practitioners and AI4SE intelligent agents working cohesively toward shared objectives as illustrated in Fig.~\ref{fig:vision}(d). This setup will involve diverse interactions: human-to-human (H-H), human-to-agent (H-A), and agent-to-agent (A-A). While substantial research exists on H-H and H-A dynamics, the A-A interactions are much less explored, especially in the software engineering field. Central to these interactions is the establishment of symbiosis and synergy, enabling mutual enrichment between practitioners and AI4SE agents.

\vspace{0.2cm}
\noindent{\em Adaptable yet solid} 
\vspace{0.2cm}

Today's software practitioners navigate a dynamic environment. Team members come in from the job market and, in time, move on. In the envisioned SE 2.0, AI4SE agents must be agile and adaptable to effortlessly integrate into teams as illustrated in Fig.~\ref{fig:vision}(e). They should be capable of discerning the strengths and capabilities of both human members and fellow AI4SE agents, identifying avenues to contribute meaningfully to the team's goals. Moreover, these agents should possess the resilience and flexibility to adjust when either their AI counterparts or software practitioners transition out of the team.

Lastly, the viability of SE 2.0 hinges on solid legal, ethical, and economic foundations as illustrated in Fig.~\ref{fig:vision}(f). There may be a need for new legal frameworks to delineate responsibility when software practitioners and intelligent agents collaborate. Privacy and copyright regulations will likely require adjustments to cater to SE 2.0 dynamics. Ethical concerns must be addressed to ensure that integrating AI4SE agents into software engineering processes yields societal benefits while mitigating potential adverse impacts, such as job losses for software practitioners. From an economic standpoint, aspects like the AI4SE agent market dynamics and vendor profitability models need attention. Therefore, the evolution and implementation of SE 2.0 will necessitate contributions not only from the Software Engineering field and Computer Science discipline but also from broader academic and professional domains.

\vspace{0.2cm}
\noindent{\em Timeline} 
\vspace{0.2cm}

During the session in which this talk was presented at ICSE 2023, an engaging discussion emerged about the timeline for SE 2.0's realization. The transition from the current SE to SE 2.0 will unfold in phases. Currently, we observe a surge of enthusiasm among software practitioners to harness AI4SE tools. However, there are mixed results and many unresolved issues; current tools are often cumbersome and ineffective for various software engineering tasks. They are also only able to automate some of the many tasks that software practitioners do today.

In the upcoming phase -- {\tt Now} to {\tt Now+U} years -- buoyed by increased AI4SE research and substantial investments from academia, industry, and government, many of the challenges mentioned in the previous paragraph will be addressed. This will transform current AI4SE tools into ``power'' tools. These ``power'' tools will still not be autonomous and require practitioners' close supervision. However, they will address and alleviate many of the frustrations practitioners currently face in using them, and be integrated smoothly across a broad range of software engineering tasks. Drawing a parallel, consider the progression of smartwatches. Their genesis can be traced back to 1976, when they were expensive, offered limited functionality like basic calculations, and had many usability issues. In contrast, the smartwatches of today, nearly half a century later, are versatile and user-friendly gadgets. Given today's accelerated pace of innovation,  it is plausible that the evolution of AI4SE tools will occur in a considerably shortened timeframe, although predicting a precise value for {\tt U} is challenging.

%to achieve this milestone, it is likely that our AI4SE journey will be much shorter, although predicting the precise value for {\tt U} is challenging.

In the subsequent phase -- {\tt Now+U} years to {\tt Now+(w$\times$U)} years -- these ``power'' tools will evolve into intelligent workmates, characterized by responsible autonomy. This progression will likely be incremental too, unfolding as the facets of responsible autonomy depicted in Fig.~\ref{fig:vision}(b) – (f) are actualized and their associated challenges addressed. It is at this juncture that Software Engineering 2.0 will truly come into being. The value for $w$ will depend on how fast Artificial General Intelligence (AGI) will be realized. Projections vary significantly, with some anticipating some form of AGI realization in a few years, while others expect it to take several decades~\cite{wsj2023deepmind,grace2022expert}.

%-- some estimated that it will be realized in a few years, yet others estimated a few decades~\cite{wsj2023deepmind,grace2022expert}.

\section{Roadmap to Trustworthy AI4SE}\label{sec:roadmap_trust}
\vspace{0.1cm}``{\em Trust but verify}'' --  Ronald Reagan \vspace{0.2cm}

\begin{figure*}[t]
    \centering
    \includegraphics[width=0.9\textwidth]{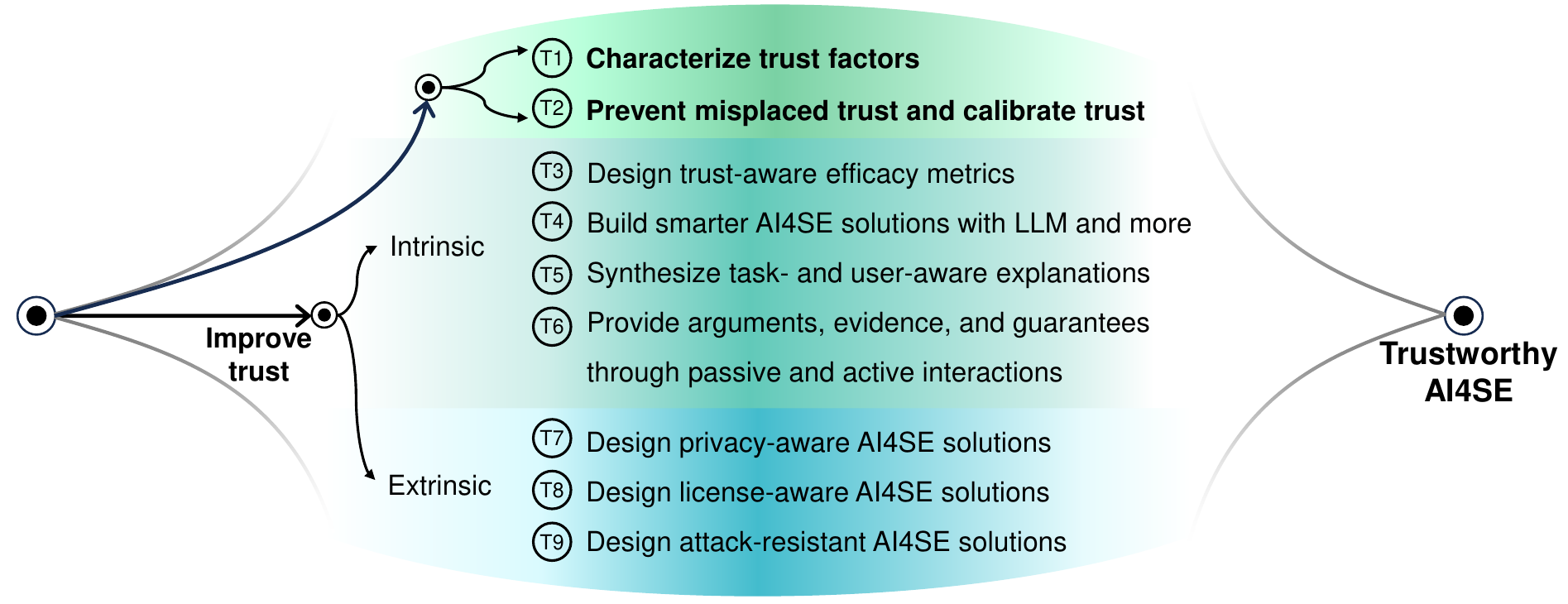}
    \caption{Roadmap to Trustworthy AI4SE}
    \label{fig:roadmap_trust}
\end{figure*}

Trust is a pivotal element in successful collaborations between humans and intelligent solutions~\cite{ParasuramanR97,LeeS04,Muir1987TrustBH}. Section~\ref{sec:challenges} has highlighted the challenges in establishing trust between software practitioners and AI4SE solutions. Thus, further research is important. This section outlines nine strategies for achieving trustworthy AI4SE, as depicted in Fig.~\ref{fig:roadmap_trust} and detailed further below.

\vspace{0.2cm}
\noindent\begin{tikzpicture}[baseline={([yshift={-0.8ex}]current bounding box.center)}]
    \draw (0,0) circle (0.3cm);
    \node at (0,0) {T1};
\end{tikzpicture}
\noindent{\em Characterize trust factors} 
\vspace{0.2cm}

Firstly, a clearer definition is required regarding the factors that influence practitioners' trust in AI4SE solutions. Some studies have looked into measuring trust in automation, e.g.,~\cite{Jian00,hoff2015trust}, however, mostly not in the context of software engineering. While some studies have delved into practitioners' expectations of specific AI4SE solutions~\cite{KochharXLL16,WanXHLYY20,HuX0WCZ22,abs-2301-03846}, the scope remains limited. A broader examination that covers many more AI4SE solutions is needed. Also, although expectation is related to trust, they are not synonymous. The distinctions between the two warrant explorations in future empirical research. 

A wide-ranging empirical study, encompassing a diverse group of software practitioners considering diverse AI4SE solutions, will aid in comprehensively understanding trust factors influencing software practitioners in working with AI4SE solutions. An exciting recent work is by Johnson et al. who interviewed 18 practitioners within and outside Microsoft and uncovered factors that influence practitioners' trust in software tools, including AI4SE ones~\cite{JohnsonBFFZ23}. There is a need for more studies in this direction to corroborate and expand upon these factors, as well as to uncover potentially context-dependent factors that align with specific software engineering tasks and characteristics of specific AI4SE solutions. 

Furthermore, it is worth noting that practitioners' perceptions of trust factors might not be entirely accurate (c.f.,~\cite{Devanbu0B16}). Therefore, a multifaceted research approach encompassing interviews, surveys, controlled experiments, and field studies is crucial. This holistic effort will pave the way to comprehensively understand ``trust'' within the AI4SE context. 

\vspace{0.2cm}
\noindent\begin{tikzpicture}[baseline={([yshift={-0.8ex}]current bounding box.center)}]
    \draw (0,0) circle (0.3cm);
    \node at (0,0) {T2};
\end{tikzpicture}
\noindent{\em Prevent misplaced trust and calibrate trust} 
\vspace{0.2cm}

Trust in automation is a nuanced matter; when rightly placed, it accelerates productivity; but when misplaced, it leads to detrimental consequences. Prior studies have highlighted the dangers of automation misuse, e.g.,~\cite{ParasuramanR97}. These concerns are also applicable to AI4SE which increasingly offers more and more automation.  Despite the advances in AI4SE, there are still frequent failures. Over time, even as AI4SE solutions become increasingly sophisticated and get closer to passing the Turing test, they are likely to still be susceptible to errors, mirroring the adage ``to err is human.'' 

Complicating this is the propensity of AI4SE solutions, especially those that are based on Large Language Models (LLMs), to yield outputs that appear fluent or accurate at first glance, but contain issues or subtle inaccuracies. To illustrate, a recent study indicated that while ChatGPT-generated answers to software engineering technical questions are almost as fluent as those of humans, they are often much less useful~\cite{abs-2307-09765}. More concerning is the risk of AI4SE solutions generating wrong or harmful outputs due to hallucination~\cite{Liu23, PearceA0DK22}. This necessitates the development of technologies that can identify the aforementioned issues, including hallucination, to prevent unwarranted trust.

Additionally, there is a need for ongoing, unbiased evaluation of AI4SE solution capabilities, especially given their rapid evolution. This ensures that both practicing and budding software practitioners can judiciously determine their trust levels. There is also a need for the design of mechanisms that allow for audit and accreditation of AI4SE solutions, either with or without a trusted third party, c.f.,~\cite{abs-2302-08500,ParkKL22}.

\vspace{0.2cm}
\noindent{\em Improve trustworthiness of AI4SE solutions}
\vspace{0.2cm}

Currently, without an exhaustive empirical study, it is challenging to pinpoint all the factors that delineate or relate to trust for diverse software engineering tasks. However, one can infer that a software practitioner's trust in an AI4SE solution might be influenced by factors that fall into two dimensions: intrinsic and extrinsic. Intrinsic factors relate to an AI4SE solution's ability to produce accurate outputs and provide explanations for its results, while extrinsic factors concern with, for example, the AI4SE solution's adherence to external authority regulations and its resilience against external threats. Sections~\ref{subsec:intrinsic} and~\ref{subsec:extrinsic} discuss some strategies to enhance the trustworthiness of AI4SE solutions in these two dimensions.

\subsection{Enhance Intrinsic Trustworthiness of AI4SE Solutions}\label{subsec:intrinsic} 

Naturally, software practitioners tend to place greater trust in AI4SE solutions that consistently deliver accurate results for assigned tasks~\cite{XiaBLL16,KochharXLL16}. Software practitioners also appreciate explanations in addition to recommendations, especially when working with AI4SE solutions~\cite{KochharXLL16}. To cultivate AI4SE solutions that earn developers' trust, the following directions that can boost the efficacy of AI4SE solutions and improve their ability to provide explanations will be worth pursuing.

\vspace{0.2cm}
\noindent\begin{tikzpicture}[baseline={([yshift={-0.8ex}]current bounding box.center)}]
    \draw (0,0) circle (0.3cm);
    \node at (0,0) {T3};
\end{tikzpicture}
\noindent{\em Design trust-aware efficacy metrics}
\vspace{0.2cm}

Evaluating the trustworthiness of AI4SE solutions demands efficacy metrics that mirror factors influencing practitioner trust. For example, several studies have highlighted top-$N$ accuracy as a good efficacy metric for an AI4SE solution that produces a ranked list of recommendations~\cite{ParninO11,KochharXLL16}. This metric recognizes that practitioners often focus only on the top $N$ recommendations, with $N$ being a small number. If an AI4SE solution produces many good recommendations beyond the top-$N$, these do not positively influence practitioners' trust in the solution. Similarly, another study~\cite{HuangXL19} has proposed the number of initial false alarms (IFA) as a metric, as first impressions matter to get a software practitioner to trust an AI4SE solution. Also, in areas like defect prediction, effort-aware metrics have been proposed~\cite{MendeK10}. These metrics are more aligned with trust factors compared to their non-effort-aware counterparts. This is because practitioners may find their trust dwindling if they do not witness an increase in favorable outcomes that is commensurate with an increased effort in scrutinizing an AI4SE solution's recommendations.

%These metrics are more aligned to trust than their non-effort-aware counterparts as practitioners are likely to lose trust in an AI4SE solution if they spend more effort to inspect the solution's recommendations without getting more positive outcomes.

\vspace{0.2cm}
\noindent\begin{tikzpicture}[baseline={([yshift={-0.8ex}]current bounding box.center)}]
    \draw (0,0) circle (0.3cm);
    \node at (0,0) {T4};
\end{tikzpicture}
\noindent{\em Build smarter AI4SE solutions with LLM and more}
\vspace{0.2cm}

The area of AI4SE offers much potential for advancement. At present, a prominent research trend within the AI4SE community is the use, adaptation, and design of Large Language Models (LLMs) to automate software engineering tasks. Current studies predominantly harness these LLMs for a subset of software engineering tasks, such as code search and code summarization~\cite{ZhouH021,LuGRHSBCDJTLZSZ21}. However, software engineering is multifaceted, encompassing more than just these tasks. While recent research has begun delving into less-explored areas, e.g., managing bug reports~\cite{zhang2023cupid}, designing software architecture~\cite{AhmadW0FAM23}, building navigation aids for software engineering Q\&A sites~\cite{he2023representation}, etc., there remains significant scope for broader exploration.

Additionally, recent studies have highlighted certain limitations of LLM for software engineering (LLM4SE). For instance, these models are not robust to minor perturbations of inputs~\cite{YangSH022,MastropaoloPGCSOB23}. Also, they can be vulnerable to shifts in data distributions, like the evolution of third-party libraries and releases of new ones~\cite{abs-2305-04106}. Moreover, a prior study has also demonstrated the challenges of LLM4SE in handling data that resides in the tail-end of the distribution~\cite{ZhouKXLHL23}. Recognizing and characterizing the limitations of LLMs and devising strategies to overcome those presents compelling directions for future research.

To boost the efficacy of LLM4SE, some recent studies have focused on enhancing the inputs and outputs of LLMs and integrating them with other techniques. For instance, some studies incorporate code graphs (e.g., control flow graph,  program dependency graph, etc.) as input to LLMs~\cite{GuoRLFT0ZDSFTDC21,ShiYW0Z0ZX22}. Others transform source code to an intermediate representation (IR) that provides a more concise and uniform input to LLMs for more effective and efficient learning~\cite{GuiWZHSXSJ22,Niu24}. Moreover, other studies emphasize selecting optimal in-context examples to bolster the efficacy of LLMs~\cite{AhmedD22}. There is also an emphasis on coupling LLMs with methods like program analysis and testing to yield superior results~\cite{Liu23,ahmed2023improving,JoshiSG0VR23, FanGMRT23,abs-2307-09163,RoziereZCHSL22}. Additionally, a few studies showcase the potential of both one-off and ongoing interactions with LLMs to refine their outputs for several software engineering tasks~\cite{Liu23,abs-2304-00385}. 

%Further, certain research focuses on the selection of ideal in-context examples to boost LLM performance~\cite{AhmedD22}. There is also an emphasis on coupling LLMs with methods like program analysis and testing for enhanced outcomes~\cite{Liu23,ahmed2023improving,JoshiSG0VR23, FanGMRT23,abs-2307-09163,RoziereZCHSL22}. Additionally, a few studies highlight the promise of both singular and continuous interactions with LLMs to fine-tune their outputs~\cite{Liu23,abs-2304-00385}.

Owing to the swift progress in LLM4SE research, this concise summary does not capture its entire scope and may be quickly outdated. As such, Systematic Literature Reviews (SLRs) on LLM4SE, e.g.,~\cite{hou2023large},  can prove invaluable. The rapid evolution in LLM4SE warrants multiple SLRs to capture the latest developments and trends. Multiple SLRs can also be conducted to examine LLM4SE from various perspectives. 

\vspace{0.2cm}
\noindent\begin{tikzpicture}[baseline={([yshift={-0.8ex}]current bounding box.center)}]
    \draw (0,0) circle (0.3cm);
    \node at (0,0) {T5};
\end{tikzpicture}
\noindent{\em Synthesize task- and user-aware explanations}
\vspace{0.2cm}

Contemporary AI4SE solutions offer diverse recommendations -- including patches to fix a bug, source code to write next, third-party library to use, and so on. However, these often come without explanations. This opacity can diminish the trust software practitioners have in these suggestions. As a result, there is a need for explainable AI4SE solutions that can realize effective practitioner-AI4SE solution interactions that engender trust and bring about trusted collective intelligence. 

While there are some efforts done in this direction, e.g.,~\cite{LiuLAMC22,3585386,MahbubSR23,WidyasariPHTZ022,abs-2302-06065}, more explorations are needed. First, we can extend existing explainable AI4SE solutions to cover more software engineering tasks. Second, we can develop their capabilities to produce {\em task-specific explanations} that are tailored to suit specific software engineering tasks and contexts. Moreover, the explanations need to be {\em user-aware}; they need to consider the specific expertise and experience of a practitioner who uses an AI4SE solution. Producing effective explanations for a recommendation made by an AI4SE solution given a specific task and context for a target software practitioner is challenging for several reasons: software engineering is a complex endeavor including diverse tasks; software engineering knowledge evolves rapidly; also, software practitioners have diverse backgrounds. Thus, there is much room to innovate to tackle these challenges.

\vspace{0.2cm}
\noindent\begin{tikzpicture}[baseline={([yshift={-0.8ex}]current bounding box.center)}]
    \draw (0,0) circle (0.3cm);
    \node at (0,0) {T6};
\end{tikzpicture}
\noindent{\em Produce arguments, evidence, and guarantees through passive and active interactions}
\vspace{0.2cm}

For greater acceptance by software practitioners, AI4SE solutions must articulate convincing arguments and present pertinent evidence to practitioners, guiding their decisions regarding recommendations made by AI4SE solutions. Furthermore, evidence should not be solely obtained passively from the data present in software artifacts. Active interaction with these artifacts -- feeding input data and monitoring the ensuing outputs -- can be an effective way of producing evidence. Additionally, it will be desirable if AI4SE solutions are able to provide some guarantees of their efficacy. 

Consider an instance where an AI4SE solution suggests to a practitioner that a particular code segment is potentially vulnerable. Its {\em argument}: the code resembles $M$ code fragments on GitHub, which experienced practitioners fixed to remove vulnerabilities that match an entry in the Common Weakness Enumeration (CWE).\footnote{https://cwe.mitre.org/index.html} To bolster this argument, the AI4SE solution can produce tangible {\em evidence} by generating a test input that showcases the exploitability of the flagged code, following techniques proposed in~\cite{AvgerinosCRSWB14,IannoneNSL21,KangNLP022}. 

%o “the code resembles the code fragments modified by a commit linked in a National Vulnerability Database vulnerability report

Similarly, an AI4SE solution recommending third-party libraries~\cite{ThungLL13} may execute static and dynamic program analyses. These will result in concrete pieces of evidence, e.g., demonstrable efficiency or a small memory footprint, to support the recommendations. A worst-case execution time guarantee can also be given by leveraging static analysis~\cite{0001CRKMF14}.

\subsection{Enhance Extrinsic Trustworthiness of AI4SE Solutions}
\label{subsec:extrinsic} 

Software practitioners need guarantees that the use of AI4SE solutions will not bring them into conflict with laws. Also, they need assurances that sufficient measures are in place to protect AI4SE solutions from malicious actors. To address these concerns, which impact the trustworthiness of AI4SE solutions, it is essential to explore the following directions.

\vspace{0.2cm}
\noindent\begin{tikzpicture}[baseline={([yshift={-0.8ex}]current bounding box.center)}]
    \draw (0,0) circle (0.3cm);
    \node at (0,0) {T7};
\end{tikzpicture}
\noindent{\em Design privacy-aware AI4SE solutions}
\vspace{0.2cm}

An open avenue of research is in fortifying AI4SE solutions to ensure compliance with privacy regulations more rigorously. Prior works have provided some protection to sensitive data in some software artifacts~\cite{BudiLJL11,LuciaLJB12,PetersMGZ13}. They, however, have mainly focused on data that comes in tabular format. They have also only considered some specific software engineering tasks, e.g., testing, defect prediction, and debugging. Thus, more can be done to strengthen this protection and expand it to cover diverse software artifacts and tasks. 

Also, privacy-aware AI4SE solutions are important for the adoption of AI4SE solutions that require software practitioners to transmit proprietary code and data to third-party services. Software practitioners (and companies) need certain guarantees that proprietary code and data are not retained or seeped into the underlying AI4SE model that can potentially be leaked to other users of the service, c.f.,~\cite{abs-2308-09932,ShokriSSS17}. The use of Trusted Execution Environments~\cite{SabtAB15} and the design of suitable protocols may be one way to achieve this guarantee, c.f.,~\cite{ParkKL22}. Alternatively, model compression techniques can be used to customize AI4SE models for local deployment on servers with limited memory and processing power~\cite{shi2022compressing}. Such local deployment eliminates the need to transfer proprietary data to third-party vendors, improving privacy. Yet another possibility is to design AI4SE solutions that employ federated learning~\cite{ShanbhagC22}.

Moreover, strategies are needed to help AI4SE solutions respect the EU General Data Protection Regulation (GDPR)'s provision of ``right to be forgotten'' without necessitating extensive retraining, c.f.,~\cite{CaoY15}. The current AI4SE solutions do not readily allow the removal of specific contributions from open-source projects when the corresponding software practitioners invoke their ``right to be forgotten'' under GDPR. 

\vspace{0.2cm}
\noindent\begin{tikzpicture}[baseline={([yshift={-0.8ex}]current bounding box.center)}]
    \draw (0,0) circle (0.3cm);
    \node at (0,0) {T8};
\end{tikzpicture}
\noindent{\em Design license-aware AI4SE solutions}
\vspace{0.2cm}

In 2022, Microsoft, GitHub, and OpenAI were sued for issues related to privacy and copyright~\cite{Metz22}. Moreover, an AI-powered coding assistant can inadvertently replicate GPL v3 licensed code from GitHub, risking copyright infringements when integrated into proprietary software. Holistic integration of licensing information and constraints~\cite{Ballhausen19} into AI4SE solutions' training or fine-tuning processes can be designed to address the aforementioned problems. Concurrently, runtime checking methods can be developed to flag AI4SE recommendations that potentially infringe on licensing terms. Code clone detection methods, e.g.,~\cite{KamiyaKI02,JiangMSG07,SainiFLBL18,MehrotraAGALP22}, can be employed as part of such runtime checks. 

\vspace{0.2cm}
\noindent\begin{tikzpicture}[baseline={([yshift={-0.8ex}]current bounding box.center)}]
    \draw (0,0) circle (0.3cm);
    \node at (0,0) {T9};
\end{tikzpicture}
\noindent{\em Design attack-resistant AI4SE solutions}
\vspace{0.2cm}
 
As AI4SE solutions become integral to software engineering processes, their ability to withstand attacks from malicious actors is of high importance. Recent studies~\cite{ YangSH022, Gao0W23, JhaR23, abs-2301-02496,nguyen2023adversarial,WanZ0SX00S22,SchusterSTS21} have examined specific attack strategies, yet a broader spectrum awaits exploration. A comprehensive threat assessment is vital, paired with detection, quantification, and mitigation strategies, aiming to design and develop AI4SE solutions that are resilient to multifaceted attacks. Strengthening AI4SE defenses, devising corrective algorithms for real-time ``self-healing'', and probing the ramifications of data poisoning on software artifacts should be prioritized. Some strides have been made in this direction, but more is warranted.

\section{Roadmap to Synergistic AI4SE}\label{sec:roadmap_synergy}
\vspace{0.1cm}``{\em The whole is greater than the sum of its parts}'' -- Aristotle \vspace{0.2cm}

\begin{figure*}[t]
    \centering
    \includegraphics[width=0.9\textwidth]{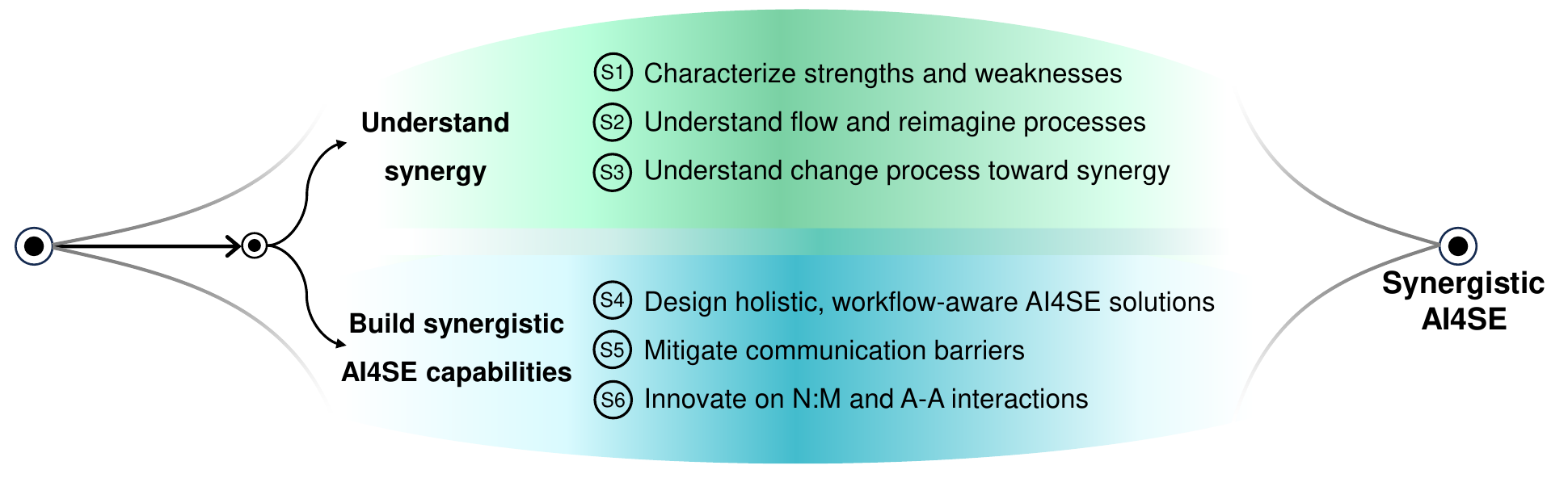}
    \caption{Roadmap to Synergistic AI4SE}
    \label{fig:roadmap_synergy}
\end{figure*}

Section~\ref{sec:challenges} highlighted a number of barriers that impede synergistic interactions between software practitioners and AI4SE solutions. Thus, further research is essential. This section outlines six strategies for achieving synergistic AI4SE, as illustrated in Fig.~\ref{fig:roadmap_synergy} and elaborated upon below.

\subsection{Understand synergy}

\vspace{0.2cm}
\noindent\begin{tikzpicture}[baseline={([yshift={-0.8ex}]current bounding box.center)}]
    \draw (0,0) circle (0.3cm);
    \node at (0,0) {S1};
\end{tikzpicture}
\noindent{\em Characterize strengths and weaknesses for better practitioner-AI4SE solution synergy}
\vspace{0.2cm}

To ensure that collaborations between software practitioners and AI4SE solutions produce outcomes greater than the sum of their individual contributions, it is essential to recognize the strengths of one party that can offset the weaknesses of the other. In software engineering, tasks often form lengthy chains, with each main task comprising several micro-tasks. Software practitioners may excel in certain micro-tasks but falter in others. Similarly, AI4SE solutions may perform very well in certain micro-tasks but perform poorly in others. The sets of micro-tasks that software practitioners and AI4SE solutions excel in may be substantially different. Hence, there is a need for empirical research to determine which micro-tasks are better assigned to practitioners -- bearing in mind that the aptitude for micro-tasks can differ among individuals and can evolve with their experience. Similarly, it is important to identify the tasks best assigned to AI4SE solutions, acknowledging that these too can vary based on the capabilities of specific AI4SE solutions and the characteristics of practitioners using them.

Also, not in all situations AI4SE solutions can help. To underscore this point, a past study at Adobe revealed that junior software practitioners have a lower adoption threshold for a bug localization solution\footnote{A bug localization solution produces a ranked list of potentially buggy source code files given a bug report.}~\cite{DBLP:journals/tse/JarmanBSTL22}. On the other hand, experienced practitioners indicated less need for such automated assistance and have a higher threshold for adoption. Therefore, for AI4SE bug localization solutions to genuinely complement these experienced software practitioners, they need to surpass a certain level of efficacy. Failing to do so, these solutions might prove more obstructive than beneficial. One possibility is for such solutions to provide recommendations sparingly only in cases for which they are likely to exceed the experienced practitioners' expectations, c.f.,~\cite{LeLL15, LeTL17, GolaghaPB20}. 

As AI4SE matures and AI4SE solutions become first-class citizens and autonomous agents, synergizing software practitioners and AI4SE agents can be conceptualized as an optimization problem. Specifically, assuming that we can quantitatively characterize the strengths and weaknesses of different software practitioners and AI4SE solutions for different tasks as weights, the problem of synergizing software practitioners and AI4SE solutions to achieve results that best exceed the sum of their individual contributions is a task assignment problem. In essence, we need to assign tasks to the most suitable software practitioners and/or AI4SE agents to maximize overall reward or efficacy.

\vspace{0.2cm}
\noindent\begin{tikzpicture}[baseline={([yshift={-0.8ex}]current bounding box.center)}]
    \draw (0,0) circle (0.3cm);
    \node at (0,0) {S2};
\end{tikzpicture}
\noindent{\em Understand flow and reimagine processes}
\vspace{0.2cm}

Software practitioners occasionally need assistance, but not incessantly. Ill-timed assistance can be counterproductive and break software practitioners' ``flow.'' While we can allow practitioners to manually toggle assistance on and off, they may not always discern the optimal moments for aid. Consequently, further studies are vital to pinpoint when AI4SE solutions are most beneficial, integrate them smoothly into the software development process to enhance software practitioners' ``flow'', tailor them to individual practitioners with varied preferences, and measure the benefits they bring to the table.

Also, many of the software engineering processes today do not prominently picture AI4SE solutions. As AI4SE solutions transition from tools to smart workmates, we may need to re-look into the existing processes and identify limitations that may impede synergy. There may be a need for new processes that facilitate symbiotic partnerships between practitioners and AI4SE agents. In such a partnership, both entities mutually benefit, working toward shared goals more effectively.

\vspace{0.2cm}
\noindent\begin{tikzpicture}[baseline={([yshift={-0.8ex}]current bounding box.center)}]
    \draw (0,0) circle (0.3cm);
    \node at (0,0) {S3};
\end{tikzpicture}
\noindent{\em Understand change process toward synergy} 
\vspace{0.2cm}

{\em Satir change model}~\cite{satir1991satir} delineates several stages that emerge when a change is introduced: previous status quo, resistance, chaos, integration, and new status quo. This model has been influential in organizational behavior and even in the adoption of new software engineering methodologies. For instance, Lindstrom and Beck noted that following the Satir change model, adoption of eXtreme Programming does not guarantee instant positive outcomes and benefits may appear gradually~\cite{lindstrom2003gets}. 

Similarly, the introduction of AI4SE solutions may require individuals to undergo a change process to achieve synergy. This change process may not instantly lead to improved outcomes. Departing from a familiar status quo, the introduction of an AI4SE solution -- which is perceived as an external element -- can incite resistance and even chaos. But this phase can subsequently give way to transformation and integration, ultimately establishing a new status quo. 

In AI4SE research, there have been limited studies on this change process. There is a need for such studies to answer questions such as: How can we measure the efficacy of the change process? What strategies can accelerate the change process, allowing us to reach the new status quo more rapidly? How can we increase the likelihood that the new status quo significantly surpasses the previous one in terms of quality (of the software developed) and productivity (of software practitioners)? And lastly, what automated solutions can be developed to alleviate the challenges software practitioners face during this transformation?

\subsection{Build synergistic AI4SE capabilities}

\vspace{0.2cm}
\noindent\begin{tikzpicture}[baseline={([yshift={-0.8ex}]current bounding box.center)}]
    \draw (0,0) circle (0.3cm);
    \node at (0,0) {S4};
\end{tikzpicture}
\noindent{\em Design holistic and workflow-aware AI4SE solutions}
\vspace{0.2cm}

Most AI4SE solutions are tailored for specific micro-tasks, such as code generation, code summarization, fault localization, or duplicate bug report detection. In contrast, as highlighted in Section~\ref{sec:challenges}, software practitioners often operate at a different level of abstraction, considering higher-level and broader objectives. This difference may be a barrier to synergy. 

To address this challenge, a shift toward holistic, workflow-aware AI4SE solutions can be beneficial. For example, as a start, AI4SE solutions can transfer information and insights from one micro-task to others, c.f.~\cite{TianLS12,NiuLNL23}. Additionally, they can leverage practitioners' interactions and feedback in one micro-task to improve their capabilities in others. Moreover, we can build AI4SE solutions that pivot from assisting isolated micro-tasks to improving the entire workflow, ensuring that they capture the broader picture rather than just the individual micro-tasks -- seeing the ``forest'' instead of ``trees''.

\vspace{0.2cm}
\noindent\begin{tikzpicture}[baseline={([yshift={-0.8ex}]current bounding box.center)}]
    \draw (0,0) circle (0.3cm);
    \node at (0,0) {S5};
\end{tikzpicture}
\noindent{\em Mitigate communication barriers} 
\vspace{0.2cm}

Many AI4SE solutions currently offer limited interactivity. Though solutions built on top of ChatGPT have begun bridging these interaction gaps, they still fall short of replicating the deeply collaborative experience akin to pair programming with a trusted colleague. Also, in spite of the strides made by LLM-powered AI4SE solutions, challenges persist, such as interactions that become stagnant or unproductive toward solving a software engineering task~\cite{Liu23}.

To truly capitalize on the potential of AI4SE solutions, we must enhance the communication capabilities between software practitioners and AI4SE solutions. The ideal scenario will allow practitioners to interact with AI4SE solutions as naturally and effectively as they do with their peers. Achieving seamless communication is pivotal for fostering synergy. Moreover, the means of communication should extend beyond just text and code to encompass visuals (like diagrams and sketches), gestures (captured via videos or wearables), and other modes of interaction. Advances in Foundation Models that go beyond LLMs and consider modalities such as images and videos, e.g.,~\cite{KoCCORK23}, may pave the way toward better communication between practitioners and AI4SE solutions.

%Furthermore, many AI4SE solutions exhibit limited interactivity. While recent solutions like ChatGPT have begun addressing these gaps, the interaction they offer is still a far cry from the collaborative experience of pair programming with a trusted colleague. Despite the advancements of large-language-model-powered solutions, there remain instances where dialogues become stagnant, deteriorate, or fail to progress effectively~\cite{Liu23}.Many of the existing AI4SE solutions have limited memory and personalization power. 

\vspace{0.2cm}
\noindent\begin{tikzpicture}[baseline={([yshift={-0.8ex}]current bounding box.center)}]
    \draw (0,0) circle (0.3cm);
    \node at (0,0) {S6};
\end{tikzpicture}
\noindent{\em Innovate on N:M and A-A interactions} 
\vspace{0.2cm}

Most existing AI4SE solutions focus on the {\tt 1:1} usage scenario, wherein a single software practitioner engages with one AI4SE solution. To more fully unlock the potential of AI4SE, we need to explore the {\tt N:M} usage scenario, wherein several software practitioners synergistically interact with multiple AI4SE solutions in a team. Research on this {\tt N:M} usage scenario is limited. Considering that most software engineering teams consist of more than one practitioner, it is important to explore how a team of interacting practitioners and AI4SE solutions can collaboratively and synergistically complete tasks. 

%\vspace{0.2cm}\noindent{\em Synergy between AI4SE agents} \vspace{0.2cm}

As AI4SE evolves to take a central role in software engineering, it becomes imperative to investigate the synergy among AI4SE solutions. Currently, limited research delves into how these solutions can recognize the capabilities of their counterparts, distribute tasks, and collaboratively operate by capitalizing on each other's strengths and addressing each other's weaknesses. These collaborative capabilities should be developed for both static (where collaborations occur among a predefined set of AI4SE solutions) and dynamic (where collaborations involve previously unknown AI4SE solutions) settings.

%As AI4SE matures to be first-class citizens in software engineering we need to look into the synergy between AI4SE agents. At the moment, there is little research done on how AI4SE agents uncover the capabilities of other AI4SE agents, how can they distribute work, and how can they work collaboratively leveraging the strengths and weaknesses of others. These capabilities need to be developed by considering both static (where collaborations are among a set of known AI4SE agents) and dynamic settings (where collaborations are between unknown AI4SE agents).

%In the field of AI recently, there is a proposal for langchain and toolformer where a language model may employ the power of other solutions (that may not be language models). There is a need to expand the existing solutions for better synergy among machines.

%\section{Legal and Ethical Considerations}\label{sec:legal_ethical}
%\input{fose/legal_ethical}

\section{Summary and Concluding Remarks}\label{sec:conclusion}
\vspace{0.1cm}``{\em If you want to go far, go together}'' --  African Proverb \vspace{0.2cm}

The area of AI for Software Engineering (AI4SE) has witnessed exponential growth over the past two decades. Evolving from a niche segment centered around a handful of software engineering tasks, AI4SE has grown into a key pillar within the software engineering field, underscored by the prowess of specialized AI4SE solutions across a wide range of software engineering tasks. Three distinct waves of innovation have shaped the trajectory of AI4SE: the surge of software engineering big data, the incorporation of deep learning into the design of AI4SE solutions, and, more recently, the development of AI4SE solutions based on large language models. The latter has propelled AI4SE into the limelight, spawning industry-grade solutions that have found widespread adoption.

This paper highlights two key challenges still confronting AI4SE: the need for trust and synergy. These intertwined principles are important for harnessing the full potential of AI4SE. Looking forward, trustworthy and synergistic AI4SE solutions can transform our present AI4SE tools into truly intelligent workmates, characterized by {\em responsible autonomy}. These AI4SE entities, acting as intelligent agents, will seamlessly integrate as key autonomous contributors,  performing varied roles adeptly and responsibly in software engineering workflows. Moreover, their adaptability will empower them to operate symbiotically not just alongside individuals but within larger, dynamic teams where members -- both humans and intelligent agents -- can dynamically change. This transformation, fueled by the progression of AI to Artificial General Intelligence (AGI), will mark the inception of Software Engineering 2.0. While advancements in software engineering and AI research are clearly important, the successful realization of the Software Engineering 2.0 vision also hinges on strong foundational frameworks in the legal, ethical, and economic domains.

To achieve the aforementioned vision of Software Engineering 2.0, this paper presents two roadmaps steering toward trustworthy and synergistic AI4SE. The roadmaps enumerate 15 open challenges that await further attention of the AI4SE community. Addressing these could move AI4SE ever closer to Software Engineering 2.0.

This paper seeks to motivate more to join and contribute to the AI4SE research journey. AI4SE is currently in a ``{\em Belle \'Epoque}''  mirroring the aeronautics revolution of the early 1900s. Just as the Wright brothers' groundbreaking 12-second flight in 1903 redefined the boundaries of travel, the promising advances of AI4SE hold the potential to reshape the future of software engineering. Similar to aviation post its inaugural flight, sustained collaboration and dedication over many years by numerous contributors is pivotal to fully realizing the power of AI4SE in revolutionizing the software engineering landscape.

\section*{Acknowledgments}
%\vspace{0.2cm}\noindent {\bf Acknowledgment.}

I would like to express my thanks and appreciation to the many people who provided insights and assistance:

\begin{itemize}

\item Special thanks to Xing Hu and Hoa Khanh Dam for their excellent organization of the FoSE track at ICSE 2023 and the compilation of the post-proceedings. 

\item The perspectives presented in this paper are shaped by discussions that I had with many colleagues, insightful papers that I read, and great talks that I attended. Notably, I learned valuable insights from the keynote of Margaret-Anne (Peggy) Storey at ASE 2022 titled ``From Automating Software Engineering to Empowering Developers.'' I also learned much from fellow speakers at the ICSE 2023 Future of Software Engineering (FoSE) track, particularly speakers at the ``AI \& SE and Debt'' session, including Thomas Zimmermann, Mark Harman, and Paris Avgeriou. The lively discussion at the end of the session also contributed to many points described in this paper. 

\item Parts of this paper were adapted from a successful grant proposal submitted to the National Research Foundation, Singapore under its Investigatorship Grant Call. Many colleagues provided valuable input for that proposal.

\item Ratnadira Widyasari, Kisub Kim, and Chengran Yang greatly help to enhance the aesthetics of Fig.~\ref{fig:vision},~\ref{fig:roadmap_trust}, and~\ref{fig:roadmap_synergy}. 

\item The paper was enriched by valuable feedback from many people who reviewed a preliminary draft, including (in alphabetical order): Bowen Xu, Ferdian Thung, Hong Jin Kang, Jieke Shi, Shaowei Wang, Ting Zhang, Xin Zhou, Yuan Tian, and Zhou Yang. 

\end{itemize}

The writing and editing of this paper exemplify a synergistic collaboration between humans and AI-powered tools. For example, many icons in Fig.~\ref{fig:overview} and~\ref{fig:vision} were created using Microsoft Image Creator\footnote{\url{https://www.bing.com/create}} and DreamStudio\footnote{\url{https://dreamstudio.ai}}, both of which offer generative-AI-powered text-to-image functionality. Also, the paper's punctuation, spelling, grammar, clarity, and engagement were improved using Grammarly and ChatGPT. While ACM permits such usage of Grammarly and ChatGPT without obligatory acknowledgement\footnote{\url{https://www.acm.org/publications/policies/frequently-asked-questions}}, it is acknowledged here for the sake of completeness.

Although I consider the perspectives presented in this paper to be well-informed, they are neither comprehensive nor without potential flaws. 

This research / project is supported by the National Research Foundation, under its Investigatorship Grant (NRF-NRFI08-2022-0002). Any opinions, findings and conclusions or recommendations expressed in this material are those of the author(s) and do not reflect the views of National Research Foundation, Singapore.

% \clearpage

\balance
\bibliographystyle{IEEEtran}
\bibliography{main}

\end{document}